\documentclass{appolb}
\usepackage{graphicx}

\begin{document}
\title{Commissioning of the new calorimeters of the KLOE-2 experiment}
\author{F.~Happacher,
\address{INFN, Laboratori Nazionali di Frascati, Frascati (Rm), Italy}
\\
\vspace{0.5cm}
{M.~Martini
}
\address{Universit\`a degli studi Guglielmo Marconi, Rome, Italy and \\ INFN, Laboratori Nazionali di Frascati, Frascati (Rm), Italy}
\\
\vspace{0.5cm}
{\it on behalf of the KLOE-2 Collaboration}
}

\maketitle
\begin{abstract}
Three new sub-detectors have been installed on May 2013 in the KLOE
apparatus of Laboratori Nazionali di Frascati of INFN. 
Photon detection is improved by means of a small crystal
calorimeter, named CCALT, in the very forward direction and of 
a tungsten-scintillating
tile sampling device, named QCALT, instrumenting the low-beta 
quadrupoles of the 
accelerator. 
During the first DA$\phi$NE operations, some preliminary runs, both with and without 
collisions, have been acquired allowing the commissioning of new subdetectors. 
In this paper, we report a brief description of QCALT and CCALT and a summary of the commissioning 
phase. 

\end{abstract}
\PACS{PACS numbers come here}
  
\section{QCALT}

In the old IP scheme of DA$\Phi$NE, the inner focalizing quadrupoles had two
surrounding calorimeters, named QCAL~\cite{qcal}, covering a polar angle
down to 21 degrees. Each calorimeter consisted of 16 azimuthal sectors
composed by alternating layers of 2 mm lead and 1 mm BC408 scintillator tiles,
for a total thickness of $\sim 5 $X$_0$. The back bending fiber arrangement
allowed the measurement of the longitudinal coordinate by time difference
with a resolution of 13 cm. These calorimeters were characterized by a low light
response (1-3 pe/mip/tile) due to the optical  coupling in air between fivers and PMT, to the fiber length
($\sim$2 m for each tile) and to the quantum efficiency of the used photomultipliers
(standard bialkali with $\sim$20\% QE). 

The presence of a calorimeter in the quadrupoles region is needed to intercept photons 
coming from $K_L$ decays and lost in the  beam
pipe. Increased efficiency for these photons  implies a 
strong reduction of the $K_L^0\to3\pi^0$ background in CP violating events
such as $K_L\to\pi^0\pi^0$. 
Moreover, high granularity in the calorimeter will help on reducing 
accidental contribution from machine background ~\cite{3pi0}.

For the  KLOE-2 upgrade~\cite{let,het,kloeIT} we have designed and built two new calorimeters,
named QCALT~\cite{qcalt_paper}. 
The detector requirements can be summarised as: 
a realization of a high granularity calorimeters with sufficient $X_0$, fast timing response with a time resolution better than $< 1\;ns$ plus some boundary conditions arising from the location of the existing detector: maximum length below 1 meter,
maximum height 5 cm and photosensors working in B field.

\subsection{QCALT design and components}

\begin{figure}[htb]
\includegraphics[width=9.5cm]{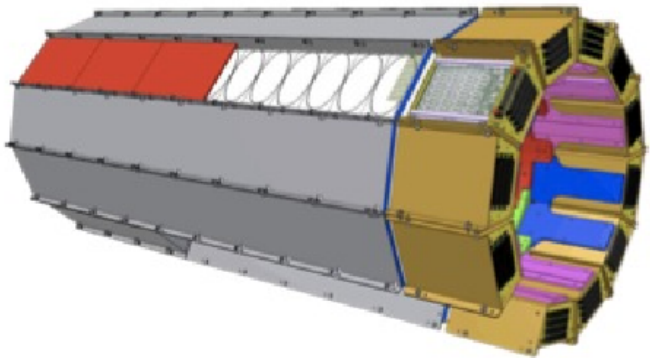}
\caption{CAD drawings of the QCALT calorimeter. Details of the tile fiber arrangement are shown as well s the FEE readout boxes.}
\label{qcalcad}
\end{figure}

QCALT consists of  two dodecagonal sampling calorimeters covering 
the quadrupoles region close to the IP. Each module, 
1/12 of one calorimeter, is composed of16 towers with 5 layers of 3.5 mm absorber
and 5 scintillator tiles 5 mm thick (See Fig.\ref{qcalcad}). 
The sampling absorber is made with a 90\% tungsten and 10\% 
copper alloy. 
Scintillator tiles are made using EJ-200 plastic scintillator produced
by Eljen and painted using BC-620 refractive paint. 
Light from scintillator
is routed outside via multi-cladding BCF-92 1mm round fibers 
inserted in a circular groove ensuring a fast emission time (5 ns/pe), long attenuation length and 
a large light yield with respect to standard single cladding fibers.

At one side of the calorimeter, fibers are collected and glued 
into a plastic holder polished with 0.05 mm planarity 
and coupled to an aluminum dioxide PCB where  80 SMD SiPM are bounded. 
Each tile is independently coupled to a photodetector
ensuring 0.05 mm clearance in each direction. 

\subsection{QCALT photosensors and FEE}

In the  QCALT we use a  custom   SMD circular SiPM 
produced by FBK of 1.2 mm diameter,
25 $\mu$m pixel size and PDE (photon detection efficiency)
peaked at 480 nm. The 80 photodetectors for each module
are bonded on an Alluminum PCB and covered with optical resin. 
The bias voltage required is between 30 and 35 V. 

Each PCB is then connected to 4 boards, 20 channels/each,
including $\times$20 trans-impedance preamplifiers, 
HV regulators (0.1\% precision, 0.01\% stability) and 
single threshold discriminators with differential output.
Each module requires a power consumption of 2 W. Boards
are inserted in a brass Faraday cage with air cooling.  Calorimeter reconstruction is  based only on timing using multi-hit TDC with  0.5 ns 
resolution. 

\subsection{QCALT commissioning}

The QCALT calorimeters have been installed on May 2013 while the  FEE boards have been
tested during each installation steps. At the end of the beam pipe insertion, we had less
than 10 dead channels over about 1800 readout channels. 

During first beam insertion, we have experienced heating problems due to insufficient 
cooling of the electronics. This problem has been solved in October 2013 with the 
installation of a dedicated air compressed line for both IT and QCALT cooling which 
decreases the SiPM temperature from 50$^\circ$ to 30$^\circ$ permitting normal operation. 

A dedicated slow control for QCALT has been developed  that allows single channel 
HV and threshold setting and temperature control.

Some preliminary runs, both with and without beam, have been collected with QCALT 
to test the complete FEE chain and SiPM functionality. Offline architecture has been developed permitting
reconstruction of QCALT data also with other subdetectors.

Monte Carlo simulation of the entire QCALT has also been completed  and is being integrated in the official 
KLOE-2 simulation framework. 

During November 2014, as soon as DA$\phi$NE will resume  operations, collision runs will be acquired to test the full functionality of the detector in collisions.

\newpage
\section{CCALT}

\begin{figure}
\begin{center}
\includegraphics[width=2.in]{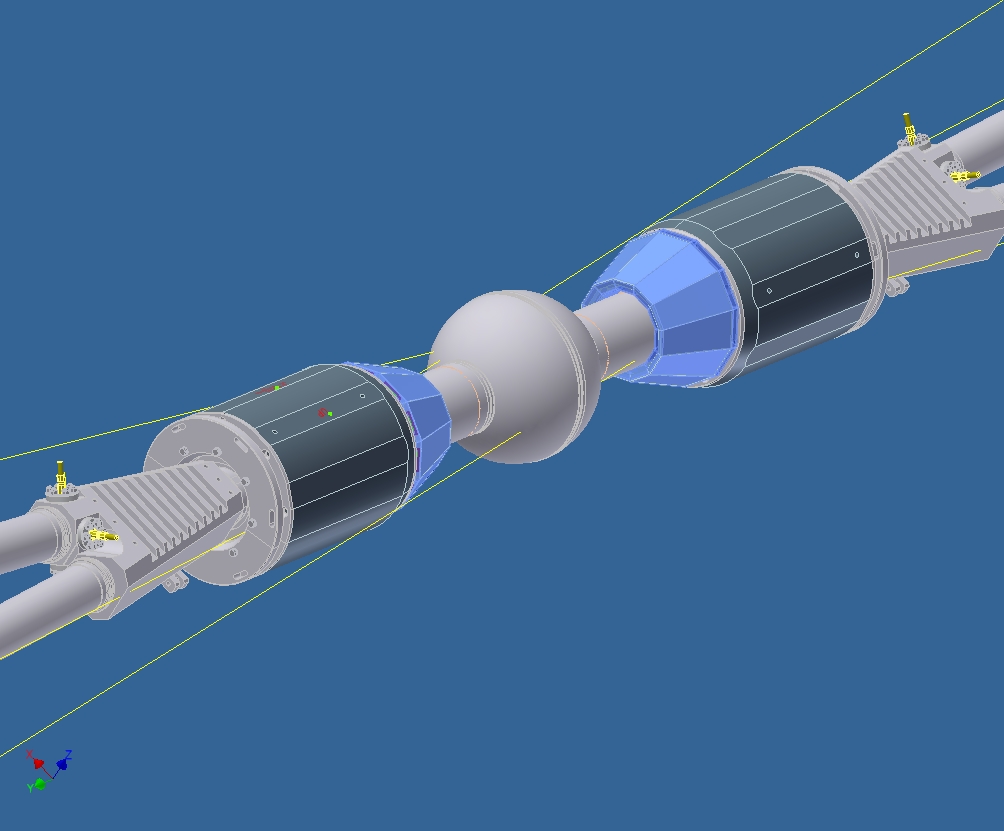} 
\caption{Zoomed-view of the IP region. The volume occupied
by the projective shaped CCALT calorimeter lies between the inner beryllium sphere and the closest
quadrupoles.}
\label{ccalt_region}
\end{center}
\end{figure}

In the upgraded KLOE-2  
detector we have introduced the CCALT~\cite{ccalt} 
calorimeter between the interaction point, IP, and the first
inner quadrupoles (See Fig.~\ref{ccalt_region}) to extend the angular 
coverage of the main electromagnetic calorimeter.   The angular coverage extension,
from a polar angle of 20$^{\circ}$ down to $8^{\circ}$, will
increase the multiphoton detection capability of the experiment 
enhancing the search reach 
for rare eta and kaon decays reconstruction, such as 
$K_S\to\gamma\gamma$, $\eta\to\pi^0\gamma\gamma$, $K_S\to\pi^0\pi^0\pi^0$ 
and  $\eta'$ prompt 
decay channels~\cite{KLOE2}.

The basic layout of the calorimeter extension
consists of two small barrels of LYSO crystals
readout with silicon photosensors; aiming to achieve a timing
resolution between 300 and 500 ps for 20 MeV photons. 
The first test of a (5.5$\times$6$\times$13) cm$^3$ prototype for such
a detector  was carried out 
at the Beam Test Facility
of Laboratori Nazionali di Frascati of INFN. 



\subsection{CCALT: a Crystal Calorimeter with Time}


The discussion of the previous section indicates that this
calorimeter has to be very dense, with a small value of
radiation length, $X_0$,  and Moliere radius, $R_m$,
not hygroscopic and with a large light output to
improve photon detection efficiency at low energy
(from 20 to 500 MeV). Moreover, the calorimeter has 
to be extremely fast in order to allow for prompt photon 
reconstruction.
Preliminary simulation studies indicates the need
to reach a time resolution of 300$\div$ 500 ps for 
20 MeV photons.

\begin{figure}
\begin{center}
\includegraphics[width=2.in, height=1.34in]{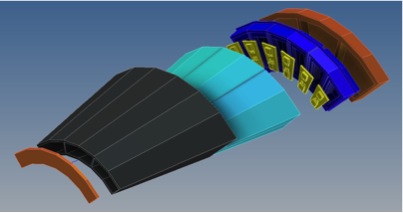} 
\includegraphics[width=2.in]{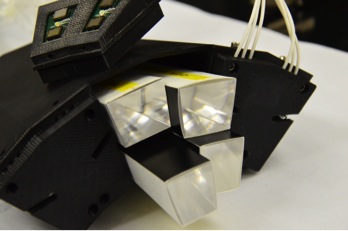} 
\caption{Drawing of an exploded view of the shell composition and actual components .}
\label{ccalt1}
\end{center}
\end{figure}

\begin{figure}
\begin{center}
\includegraphics[width=2.in, height=1.8in]{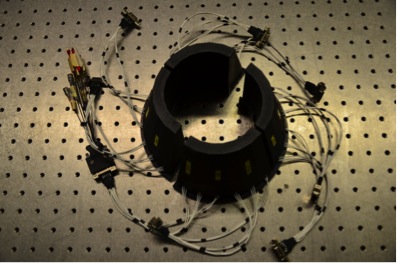} 
\includegraphics[width=2.in, height=1.8in]{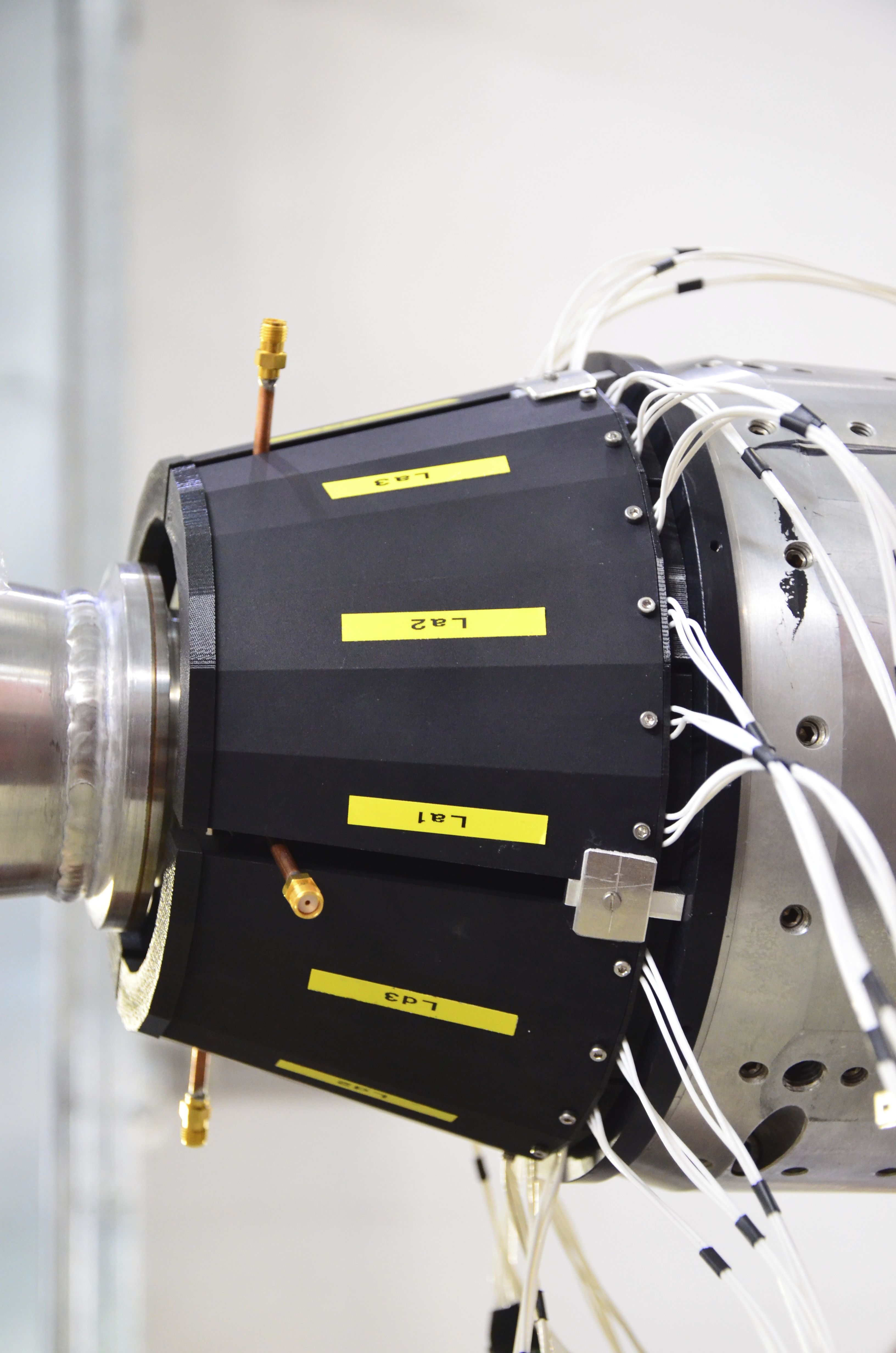} 
\caption{View of the assembled shells of one CCALT side and its mounting on the Beam Pipe .}
\label{ccalt2}
\end{center}
\end{figure}

The detector layout consists of 
two concentrical barrels composed by 48 crystals each with transversal 
dimensions of 1.5$\times$2 cm$^2$ in the readout plane,  $  0.5\times 1.5 $ mm$^2$ in the front side and longitudinal length between 13 and 15 cm.
The best crystal choice matching the requirements is provided by 
LYSO, which has  $X_0$ ($R_M$) value of 
1.1   (2 cm) and a scintillation emission time 
$\tau_{LYSO} $ of 40  ns.
Each side of the CCALT is composed by four shells made of aluminum, each constituted by three sectors containing a module with a granularity of 4 crystals.
The shells have a projective shape from the interaction point outward. Each shell has been machined using  electro discharge machining. The crystals building up a module are of two geometrical shapes
(UP and DOWN type) and keep the projective geometry.
An exploded view  of the components of each shell is shown in Figure ~\ref{ccalt1}. 
The crystals are wrapped using a 3M super reflective tape, each pair of crystals is faced to a FEE board housing two SiPM and a LED for calibrating the photo device gain.
In Figure ~\ref{ccalt2} we show the shells composing half of the CCCALT (left) and its positioning over the beam pipe (right).

In the final location of the CCALT inside KLOE-2,
the presence of an axial magnetic field of 0.52 kGauss forced
the usage of silicon based  photodetectors. 
We have used  large area Advansid  SiPM, with an active area of $4\times4\;$ mm$^2$.
The timing and energy resolution properties 
 of a LYSO crystal calorimeter readout with Silicon photosensors have
been measured building a  medium size crystal matrix prototype 
with transversal radius larger than 2 $R_m$,  longitudinal 
dimensions  between 
13 and 15 cm (corresponding to 11 $\div$ 12 $X_{0}$ of 
longitudinal containment).
The prototype consisted of an inner matrix of 10 LYSO crystals
readout by APD and an outer matrix, for leakage recovery.
Each crystal was wrapped with 100 $\mu$m of tyvek on the lateral faces,
leaving free both the front and end faces, thus allowing
one to bring calibration light pulses through an external 
LED and a fast change of the photosensors readout. 


\begin{figure}
\begin{center}
\includegraphics[width=2.1in]{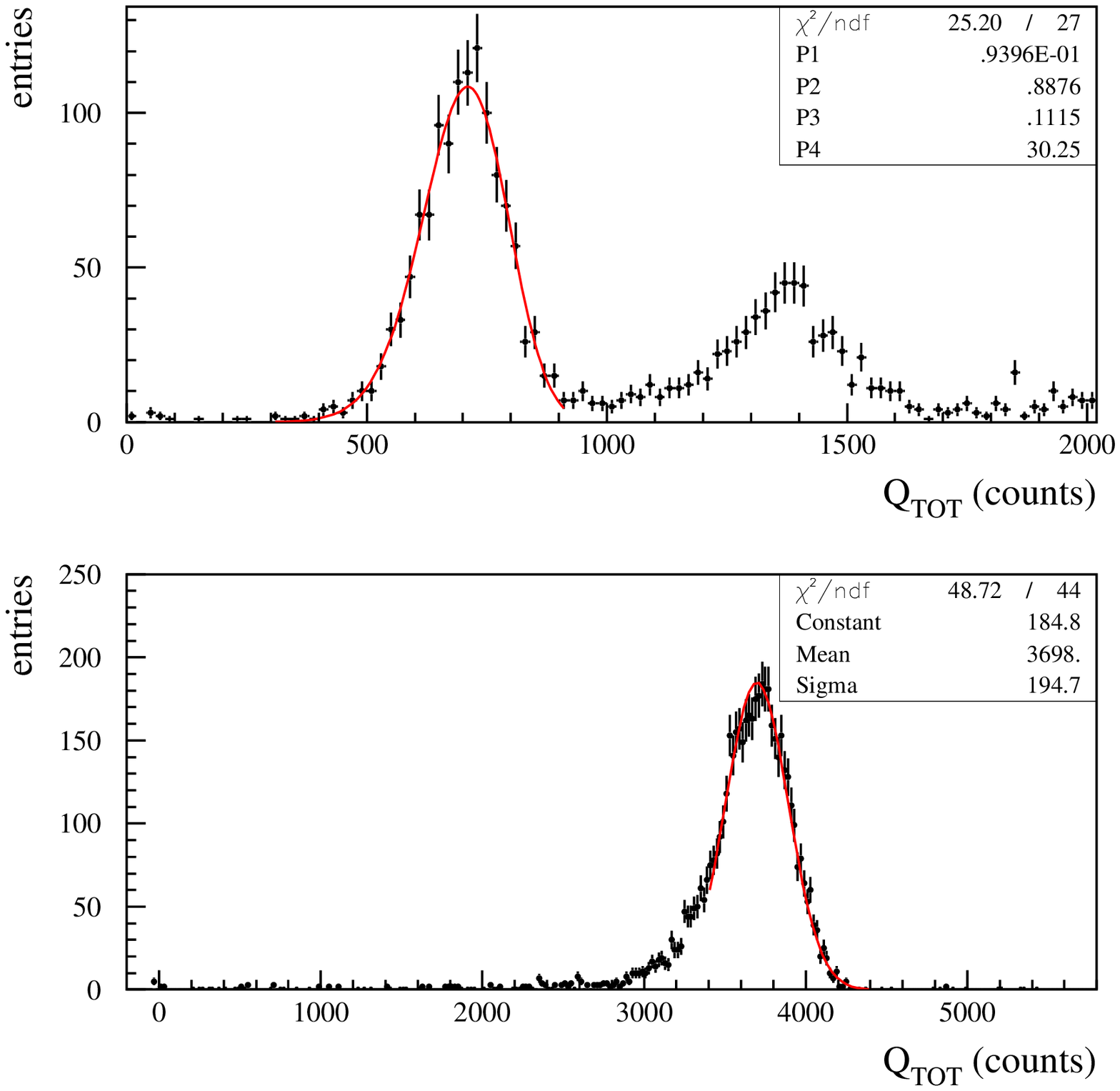}
\includegraphics[width=2.1in]{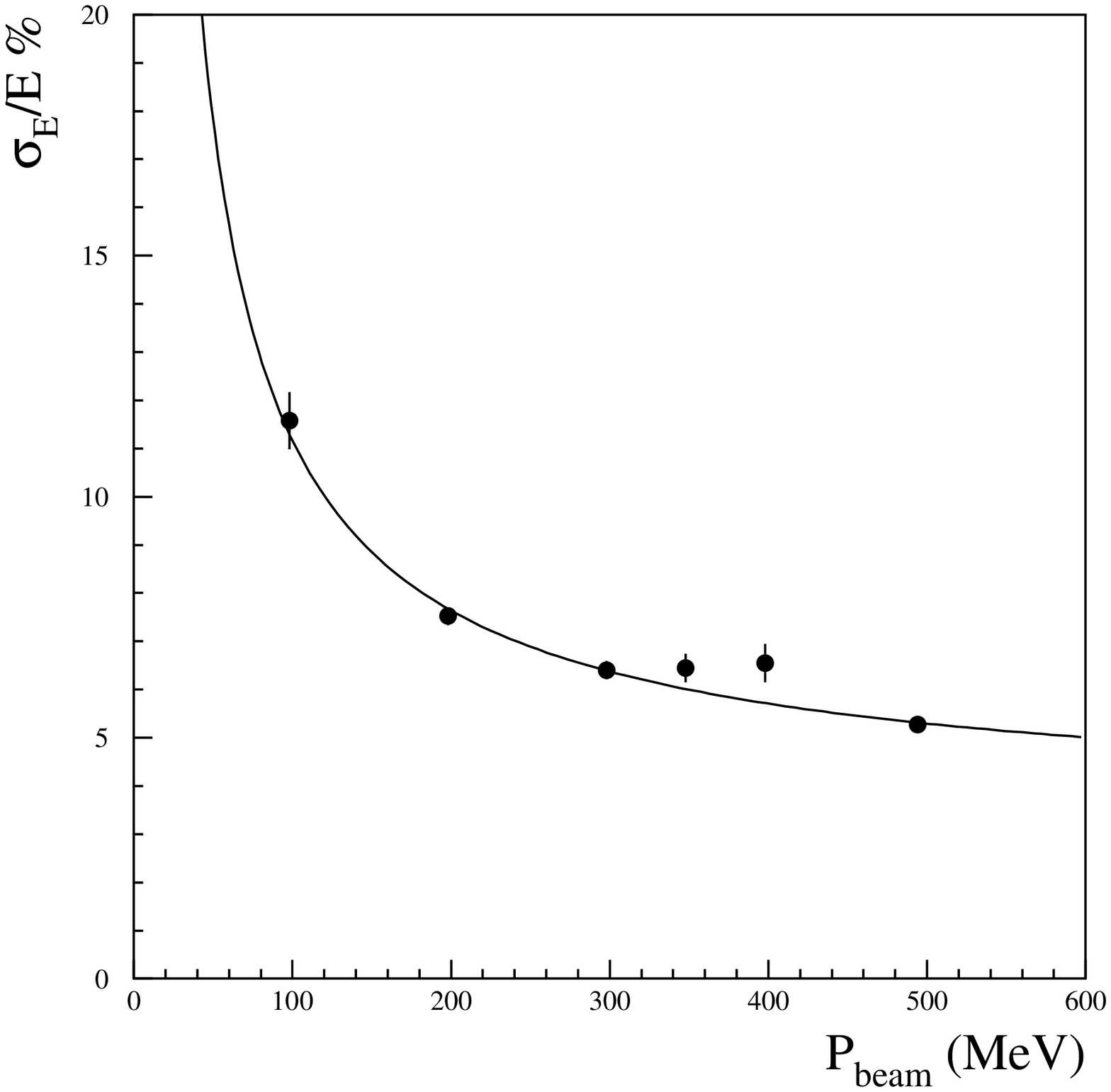}
\caption{(Left) Distribution of $Q_{TOT}$ for single electron
events at 100 MeV (top) with a logG fit superimposed
and at  500 MeV (bottom)  with a gaussian fit.
(Right) Dependence of the energy resolution on beam
momentum}
\label{energy_response}
\end{center}
\end{figure}

We have taken data at the Beam Test Facility, BTF, of LNF for
two weeks in april 2009.
In Fig.~\ref{energy_response}.left,
we show the distribution of $Q_{tot}$ for a beam of 100
and 500 MeV respectively.
In Fig.~\ref{energy_response} (right), we show the energy dependence
of the energy resolution measured on data which has been
fit with the following equation:
$\sigma_E/E = a \oplus {\rm b/(E/GeV)} \oplus \rm{c}/\sqrt{\rm E/GeV}$,
where, accordingly to MC, we have fixed the constant term to be 5 \%. 
We found $b=1.1\%$ and $c=1.4\%$ when using the gaussian fits to
the spectra. We have then determined the position resolution 
by using the BPM of BTF and  
we observed a position resolution of 2.8 $\div$ 3 mm at 500 MeV.

The calorimeter timing performances have been measured after correcting, 
event by event, for the arrival time of the electrons in the LINAC spill.
This was done by measuring the timing with the scintillators. 
The weighted energy average over all calorimeter, $T_{clu}$,  was done
after subtracting the average T$_0$ of each cell. 
A clean gaussian response is observed with
a time resolution, $\sigma_T$, of  $\sim$ 49 ps 
($\sim$ 120 ps) at 500 MeV (100 MeV) after correcting 
for trigger jitter.
\subsection{Test of single crystals with SiPM read-out}

Due to space constraints and thermal considerations, the final choice
for the read-out is large area (($4\times 4$) mm$^2$) SiPM from ADVANSID.
This results in a loss of $\sim 4$ in time resolution, of which 2.5 due
to the area reduction and 1.6 due to the lower quantum efficiency. The 
usage of SiPM still satisfies detector requirements, adding the possibility 
to increase read-out granularity due to the lower cost with respect to
APD.

\begin{figure}[!t]
\centering
\includegraphics[width=2.8in, height=2.25in]{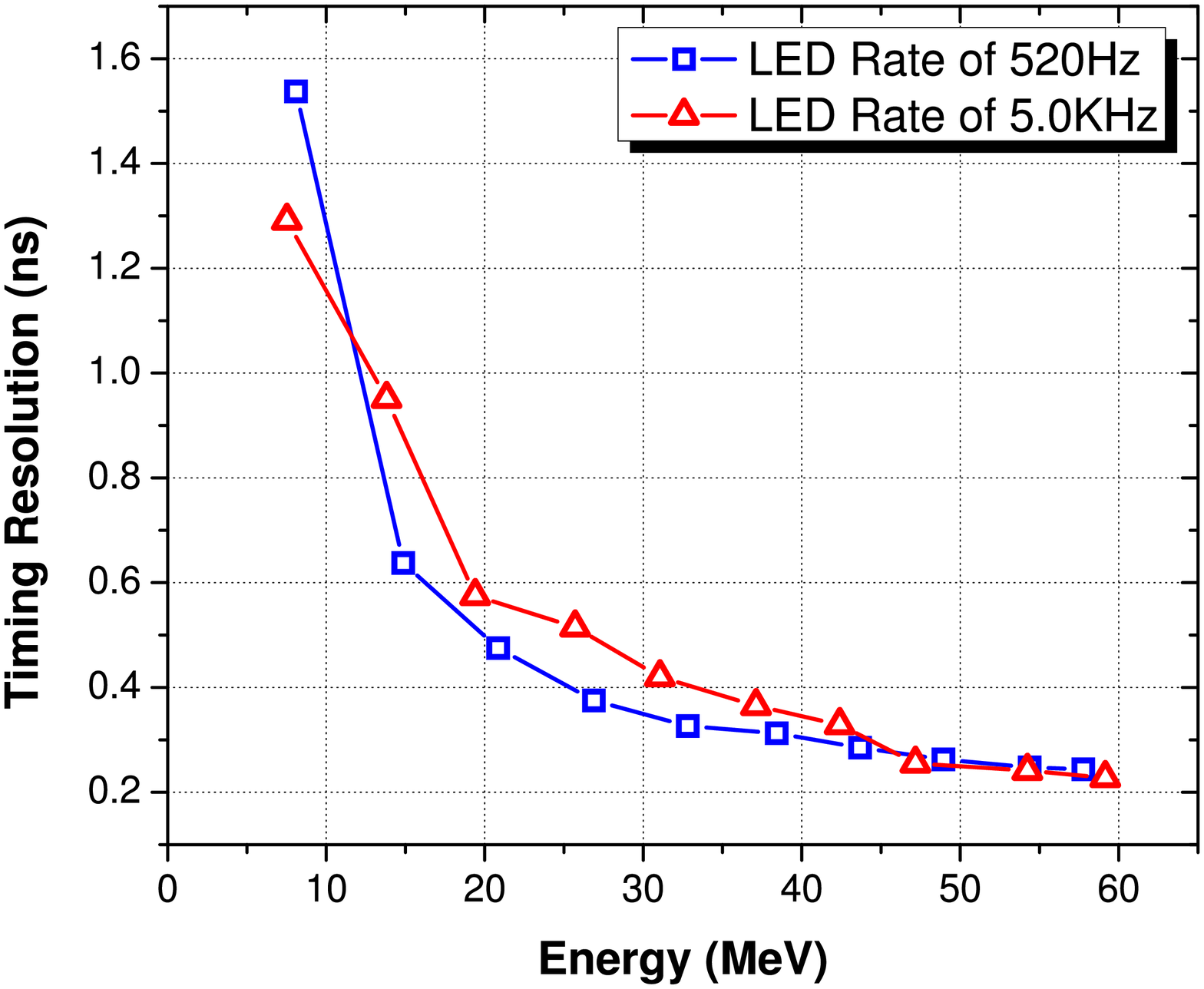}
\includegraphics[width=2.1in]{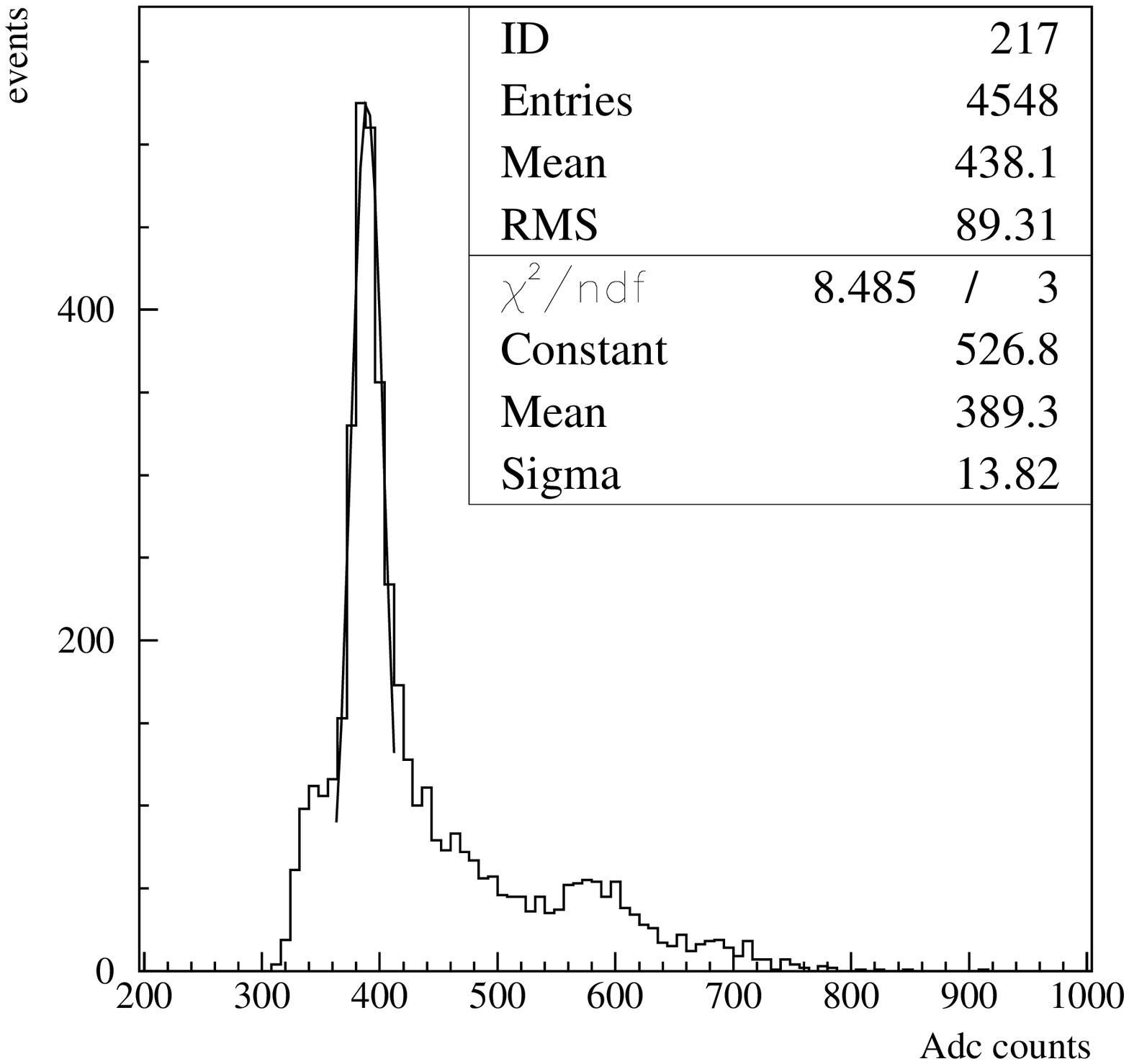}
\caption{Time resolution for a single LYSO crystal read-out by large
  area SiPM. The two curves are obtained with different LED rates (left).
  Energy distribution for $^{22}$Na source, using a single crystal 
  of the CCALT calorimeters (right).}
\label{Fig:SiPMtest}
\end{figure}

We have tested the timing performances and the dependence of the response 
on rate firing single crystals of ($20\times 20\times 150$) mm$^2$ with a 
UV LED and reading them out with the ADVANSID SiPMs. Both time resolution 
and energy response are stable for different LED rates, up to 100-200 kHz.
In Fig.~\ref{Fig:SiPMtest}, the achieved time resolution is shown as a 
function of the equivalent photon energy for 0.5 and 5 kHz LED rate.
The energy 
distribution for one crystal is reported in Fig.~\ref{Fig:SiPMtest}.
A 10\% energy resolution in obtained at 511 keV.

\begin{figure}[!t]
\centering
\includegraphics[width=2.1in,height=1.5in]{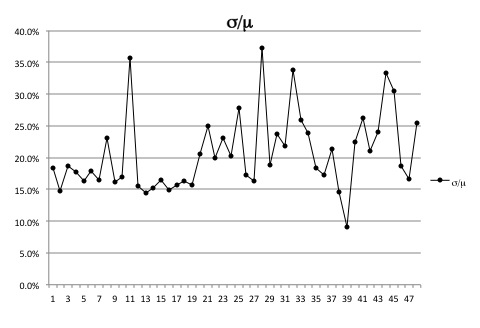}
\includegraphics[width=2.1in,height=1.5in]{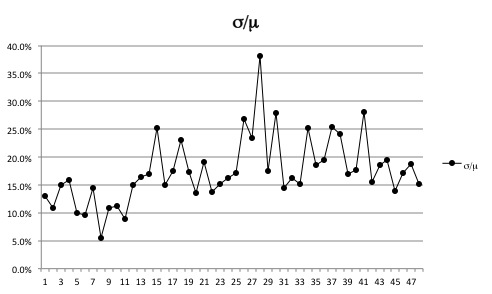}
\caption{Energy resolution distribution measured using a  $^{22}$Na source 
for UP (left) and DOWN (right) type crystals.}
\label{Fig:test}
\end{figure}

Prior to the assembling the calorimeter, all the crystals 
purchased  from SICCAS have been tested 
using a $^{22}$Na source and PM read-out. The measured energy resolution is shown in Fig.~\ref{Fig:test}; we rejected, and asked for replacement, crystals with a energy resolution $\ge 30\%$.


\subsection{Commissioning}
The CCALT detector is installed over the beam pipe,
 the slow control for HV settings (0.1$\%$ precision and 0.01\% stability of the SiPM) is working.
 All the channels are working and are being pulsed with LED and calibrated with cosmic runs.
 The CCALT channels have been included in the data acquisition and in the Offline chain of KLOE-2.

\end{document}